\documentclass[aps,prl,twocolumn,superscriptaddress,showpacs,multicol]{revtex4}
\usepackage{amsmath}
\usepackage[dvips]{graphicx}
\usepackage{bm}
\usepackage{dcolumn}

\begin{document}

\vskip 0.1truein
\title{
Cross sections and Rosenbluth separations \\
in $\rm{^1H(e,e^\prime K^+)\Lambda }$ up to $Q^2$=2.35 GeV$^2$}

\author{M.~Coman}
\affiliation{Florida International University, Miami, Florida 33199, USA}

\author{P.~Markowitz}
\affiliation{Florida International University, Miami, Florida 33199, USA}

\author{K.A.~Aniol}
\affiliation{ California State University, Los Angeles, Los Angeles, 
California 90032, USA}

\author{K.~Baker}
\affiliation{Hampton University, Hampton, Virginia 23668, USA}  

\author{W.U.~Boeglin}
\affiliation{University of Maryland, College Park, Maryland 20742, USA}    

\author{H.~Breuer}
\affiliation{University of Maryland, College Park, Maryland 20742, USA}    

\author{P.~Byd\v{z}ovsk\'y}
\affiliation{Nuclear Physics Institute, \v{R}e\v{z} near Prague, Czech
Republic}

\author{A.~Camsonne}
\affiliation{Universit\'{e} Blaise Pascal/IN2P3, F-63177 Aubi\`{e}re, France}

\author{J.~Cha}
\affiliation{Hampton University, Hampton, Virginia 23668, USA}  

\author{C.C.~Chang}
\affiliation{University of Maryland, College Park, Maryland 20742, USA}    

\author{N.~Chant}
\affiliation{University of Maryland, College Park, Maryland 20742, USA}    

\author{J.-P.~Chen}
\affiliation{Thomas Jefferson National Accelerator Facility, Newport News, 
Virginia 23606, USA}

\author{E.A.~Chudakov}
\affiliation{Thomas Jefferson National Accelerator Facility, Newport News, 
Virginia 23606, USA}

\author{E.~Cisbani}
\affiliation{Istituto Nazionale di Fisica Nucleare, Sezione di Roma1, 
Piazza A. Moro, Rome, Italy}

\author{L.~Cole}
\affiliation{Hampton University, Hampton, Virginia 23668, USA}

\author{F.~Cusanno}                                                   
\affiliation{Istituto Nazionale di Fisica Nucleare, Sezione di Roma1, 
Piazza A. Moro, Rome, Italy}

\author{C.W.~de~Jager}
\affiliation{Thomas Jefferson National Accelerator Facility, Newport News, 
Virginia 23606, USA}

\author{R.~De~Leo}
\affiliation{Istituto Nazionale di Fisica Nucleare, Sezione di Bari and 
University of Bari, I-70126 Bari, Italy} 

\author{A.P.~Deur}
\affiliation{University of Virginia, Charlottesville, Virginia 22904, USA}

\author{S.~Dieterich}
\affiliation{Rutgers, The State University of New Jersey, Piscataway, 
New Jersey 08855, USA}  

\author{F.~Dohrmann}
\affiliation{Argonne National Laboratory, Argonne, Illinois 60439, USA}

\author{D.~Dutta}
\affiliation{Massachussets Institute of Technology, Cambridge, Massachusetts 
02139, USA}

\author{R.~Ent}
\affiliation{Thomas Jefferson National Accelerator Facility, Newport News, 
Virginia 23606, USA}

\author{O.~Filoti}
\affiliation{Istituto Nazionale di Fisica Nucleare, Sezione di Roma1, gruppo 
collegato  Sanit\`a, and Istituto Superiore di Sanit\'a, I-00161 Roma, Italy}

\author{K.~Fissum}
\affiliation{Massachussets Institute of Technology, Cambridge, Massachusetts 
02139, USA}

\author{S.~Frullani}                                                   
\affiliation{Istituto Nazionale di Fisica Nucleare, Sezione di Roma1, gruppo 
collegato  Sanit\`a, and Istituto Superiore di Sanit\'a, I-00161 Roma, Italy}

\author{F.~Garibaldi}
\affiliation{Istituto Nazionale di Fisica Nucleare, Sezione di Roma1, gruppo 
collegato  Sanit\`a, and Istituto Superiore di Sanit\'a, I-00161 Roma, Italy}

\author{O.~Gayou}
\affiliation{Massachussets Institute of Technology, Cambridge, Massachusetts 
02139, USA}

\author{F.~Gilman}           
\affiliation{Rutgers, The State University of New Jersey, Piscataway, 
New Jersey 08855, USA}  

\author{J.~Gomez}
\affiliation{Thomas Jefferson National Accelerator Facility, Newport News, 
Virginia 23606, USA}

\author{P.~Gueye}
\affiliation{Hampton University, Hampton, Virginia 23668, USA}  

\author{J.O.~Hansen}
\affiliation{Thomas Jefferson National Accelerator Facility, Newport News, 
Virginia 23606, USA}

\author{D.W.~Higinbotham}
\affiliation{Thomas Jefferson National Accelerator Facility, Newport News, 
Virginia 23606, USA} 

\author{W.~Hinton}
\affiliation{Hampton University, Hampton, Virginia 23668, USA}  

\author{T.~Horn}
\affiliation{University of Maryland, College Park, Maryland 20742, USA}    

\author{B.~Hu}

\author{G.M.~Huber}
\affiliation{University of Regina, Regina, SK, S4S-OA2, Canada}

\author{M.~Iodice}                                                   
\affiliation{Istituto Nazionale di Fisica Nucleare, Sezione di Roma Tre, 
I-00146 Roma, Italy}

\author{C.~Jackson}
\affiliation{Hampton University, Hampton, Virginia 23668, USA}  

\author{X.~Jiang}
\affiliation{Rutgers, The State University of New Jersey, Piscataway, 
New Jersey 08855, USA}  

\author{M.~Jones}
\affiliation{The College of William and Mary in Virginia, Williamsburg, Viginia
23187, USA}

\author{K.~Kanda}
\affiliation{Tohoku University, Sendai, 980-8578, Japan}

\author{C.~Keppel}
\affiliation{Hampton University, Hampton, Virginia 23668, USA}  

\author{P.~King}
\affiliation{University of Maryland, College Park, Maryland 20742, USA}    

\author{F.~Klein}
\affiliation{Florida International University, Miami Florida, 33199}

\author{K.~Kozlov}
\affiliation{University of Regina, Regina, SK, S4S-OA2, Canada}

\author{K.~Kramer}
\affiliation{The College of William and Mary in Virginia, Williamsburg, Viginia
23187, USA}

\author{L.~Kramer}
\affiliation{Florida International University, Miami Florida, 33199}

\author{L.~Lagamba}
\affiliation{Istituto Nazionale di Fisica Nucleare, Sezione di Bari and 
University of Bari, I-70126 Bari, Italy} 

\author{J.J.~LeRose}
\affiliation{Thomas Jefferson National Accelerator Facility, Newport News, 
Virginia 23606, USA}

\author{N.~Liyanage}
\affiliation{Thomas Jefferson National Accelerator Facility, Newport News, 
Virginia 23606, USA}

\author{D.J.~Margaziotis}
\affiliation{ California State University, Los Angeles, Los Angeles, 
California 90032, USA}

\author{S.~Marrone}
\affiliation{Istituto Nazionale di Fisica Nucleare, Sezione di Bari and 
University of Bari, I-70126 Bari, Italy} 

\author{K.~McCormick}
\affiliation{Rutgers, The State University of New Jersey, Piscataway, 
New Jersey 08855, USA}  

\author{R.W.~Michaels}
\affiliation{Thomas Jefferson National Accelerator Facility, Newport News, 
Virginia 23606, USA}

\author{J.~Mitchell}
\affiliation{Thomas Jefferson National Accelerator Facility, Newport News, 
Virginia 23606, USA}

\author{T.~Miyoshi}
\affiliation{Tohoku University, Sendai, 980-8578, Japan}

\author{S.~Nanda}
\affiliation{Thomas Jefferson National Accelerator Facility, Newport News, 
Virginia 23606, USA}

\author{M.~Palomba}
\affiliation{Istituto Nazionale di Fisica Nucleare, Sezione di Bari and 
University of Bari, I-70126 Bari, Italy} 

\author{V.~Pattichio}
\affiliation{Istituto Nazionale di Fisica Nucleare, Sezione di Bari and 
University of Bari, I-70126 Bari, Italy} 

\author{C.F.~Perdrisat}
\affiliation{College of William and Mary, Williamsburg, Virginia 23187, USA}

\author{E.~Piasetzky}
\affiliation{School of Physics and Astronomy, Sackler Faculty of Exact Science, 
Tel Aviv University, Tel Aviv 69978, Israel}

\author{V.A.~Punjabi}
\affiliation{Norfolk State University, Norfolk, Virginia 23504, USA}

\author{B.~Raue}
\affiliation{Florida International University, Miami, Florida 33199, USA}

\author{J.~Reinhold}
\affiliation{Florida International University, Miami, Florida 33199, USA}

\author{B.~Reitz}
\affiliation{Thomas Jefferson National Accelerator Facility, Newport News, 
Virginia 23606, USA}

\author{R.E.~Roche}
\affiliation{Florida State University, Tallahassee, Florida 32306, USA}

\author{P.~Roos}
\affiliation{University of Maryland, College Park, Maryland 20742, USA}    

\author{A.~Saha}
\affiliation{Thomas Jefferson National Accelerator Facility, Newport News, 
Virginia 23606, USA}

\author{A.J.~Sarty}
\affiliation{Florida State University, Tallahassee, Florida 32306, USA}

\author{Y.~Sato}
\affiliation{Tohoku University, Sendai, 980-8578, Japan}

\author{S.~\v{S}irca}
\affiliation{Massachussets Institute of Technology, Cambridge, Massachusetts 
02139, USA}

\author{M.~Sotona}
\affiliation{Nuclear Physics Institute, \v{R}e\v{z} near Prague, Czech
Republic}

\author{L.~Tang}
\affiliation{Hampton University, Hampton, Virginia 23668, USA}  

\author{H.~Ueno}
\affiliation{Yamagata University, Yamagata 990-8560, Japan}

\author{P.E.~Ulmer}
\affiliation{Old Dominion University, Norfolk, Virginia 23508, USA} 

\author{G.M.~Urciuoli}
\affiliation{Istituto Nazionale di Fisica Nucleare, Sezione di Roma1, 
Piazza A. Moro, Rome, Italy}

\author{A.~Uzzle}
\affiliation{Old Dominion University, Norfolk, Virginia 23508, USA} 

\author{A.~Vacheret}
\affiliation{Daphnia/SPhN, CEN Saclay, 91191 Gif-sur-Yvette Cedex, France}
 
\author{K.~Wang}
\affiliation{University of Virginia, Charlottesville, Virginia 22904, USA}

\author{K.~Wijesooriya}
\affiliation{Argonne National Laboratory, Argonne, Illinois 60439, USA}

\author{B.~Wojtsekhowski}
\affiliation{Thomas Jefferson National Accelerator Facility, Newport News, 
Virginia 23606, USA}

\author{S.~Wood}
\affiliation{Thomas Jefferson National Accelerator Facility, Newport News, 
Virginia 23606, USA}

\author{I.~Yaron}                                                        
\affiliation{School of Physics and Astronomy, Sackler Faculty of Exact Science, 
Tel Aviv University, Tel Aviv 69978, Israel}

\author{X.~Zheng}                                                        
\affiliation{Massachussets Institute of Technology, Cambridge, Massachusetts 
02139, USA}

\author{L.~Zhu}
\affiliation{Massachussets Institute of Technology, Cambridge, Massachusetts 
02139, USA}

\collaboration{Jefferson Lab Hall A Collaboration}
\noaffiliation

\date{\today}

\begin{abstract}
  The kaon electroproduction reaction $^1$H(e,e'K$^+$)$\Lambda$ was
  studied as a function of the virtual-photon four-momentum, $Q^2$,
  total energy, $W$, and momentum transfer, $t$, for different values
  of the virtual-photon polarization parameter. Data were taken at
  electron beam energies ranging from 3.40 to 5.75 GeV.  The center of
  mass cross section was determined for 21 kinematics corresponding
  to $Q^2$ of 1.90 and 2.35 GeV$^2$ and the longitudinal, $\sigma_L$,
  and transverse, $\sigma_T$, cross sections were separated using the
  Rosenbluth technique at fixed $W$ and $t$.  The separated cross
  sections reveal a flat energy dependence
  at forward kaon angles not satisfactorily described by
  existing electroproduction models.  Influence of the kaon pole on
  the cross sections was investigated by adopting an off-shell form
  factor in the Regge model which better describes the observed energy
  dependence of $\sigma_T$ and $\sigma_L$.
\end{abstract}

\pacs{xxxxx}

\maketitle

Understanding the structure of nuclei and the interaction between
nucleons in terms of sub-nucleonic degrees of freedom (quarks and
gluons) is the goal of intermediate-energy nuclear physics.  The
advantage of electron scattering is that the one-photon exchange is a
good approximation and can be calculated precisely \cite{feynman}.
This allows factorization of the electron and hadron dynamics in the
electroproduction cross section.  It is generally accepted that at
four momentum transfers, $Q^2\ge 1$ GeV$^2$, the virtual photon probes
the sub-nucleonic structure of the hadron (see e.g. \cite{halzen}).
Electron beams in the energy range used at the Thomas Jefferson
National Accelerator Facility (JLab) therefore can access the
sub-nucleonic structure of hadrons.  However, these energies probe only
nonperturbative aspects of QCD.  In the nonperturbative region,
effective hadronic models play an essential role and experimental
testing is crucial in understanding the underlying physics of hadron
electroproduction.  The JLab program on electromagnetic strangeness
production is of particular interest due to the presence of the
additional strange flavour degree of freedom. High-precision data from
experiments on kaon electroproduction\cite{CLAS,Moh02} let current
models (e.g. the Saclay-Lyon \cite{SLA} and Regge \cite{RM} models) be
refined (fitting parameters and revising underlying assumptions).

The cross section of the exclusive $\rm{ ^1H(e,e^\prime K^+)\Lambda }$ 
reaction with unpolarized electrons can be expressed in terms of 
the \mbox{$\gamma^*+p \rightarrow K^++\Lambda$}
virtual photoproduction binary-process cross section as:
\begin{equation}\label{d5sigma}{{d^5\sigma}\over{dE_e' d\Omega_e d\Omega_K}} =
  \Gamma {{d\sigma(\gamma^*,K)} \over {d\Omega_K}}\end{equation}
\fussy\noindent $\Gamma$ being the virtual photon flux.

In turn, the center-of-mass virtual photoproduction cross section can
be expressed via four separated cross sections: 
\begin{equation}\label{dsig}{{d\sigma(\gamma^*,K)}\over{d\Omega_K}} = \sigma_T +
\epsilon\sigma_L +\epsilon\sigma_{TT}\cos2\Phi +
\sqrt{2\epsilon(\epsilon+1)} \sigma_{LT} \cos\Phi\end{equation}

\noindent where
$\epsilon = { 1/{[1+{2|\vec q|^2}/{Q^2}}\tan^2{(\vartheta_e/2)}]}$ is
the photon polarization parameter, and $\Phi$ is the angle between 
the leptonic plane (defined by the incoming and outgoing electrons), 
and reaction plane (defined by the virtual-photon and kaon 3-momenta).
 
The terms in eq.\eqref{dsig} correspond to the cross section for
transverse ($\sigma_T$), longitudinal ($\sigma_L$),
transverse-transverse interference ($\sigma_{TT}$) and
longitudinal-transverse interference ($\sigma_{LT}$) kaon production
by virtual photons. They only depend on the variables $Q^2$
($=-q^2$, the squared virtual-photon 4-momentum), $W$ (the
photon-nucleon center of mass energy) and $t$ (the squared 4-momentum
transfer to the kaon).

In ``parallel'' kinematics (the virtual-photon and kaon 3-momenta 
are parallel), the interference terms vanish, allowing the separation of
the longitudinal and transverse parts with the Rosenbluth technique.

The results from the E98-108 experiment~\cite{proposal} in Hall A at
Jefferson Lab presented in this letter provide new information on the
behavior of the separated cross sections of the
$\rm{ ^1H(e,e^\prime K^+)\Lambda}$ exclusive reaction in an unexplored region of
$Q^2$, $W$, and $t$ where no separations have been performed.

The E98-108 data were taken using the JLab continuous electron beam
with currents as high as 100 $\mu$A.  The beam was scattered off a 15
cm, cryogenic liquid hydrogen target (LH$_2$).  Background
distributions from the aluminum windows were obtained from an empty
target cell replica.  The scattered electrons and kaons were detected
in coincidence in two High Resolution Spectrometers
(HRS)\cite{nimHallA}.  The HRSs can achieve a momentum resolution of 2
$\times$ 10$^{-4}$ and an angular resolution of about 2 mrad.

\begin{figure}
\resizebox{0.45\textwidth}{!}{\includegraphics{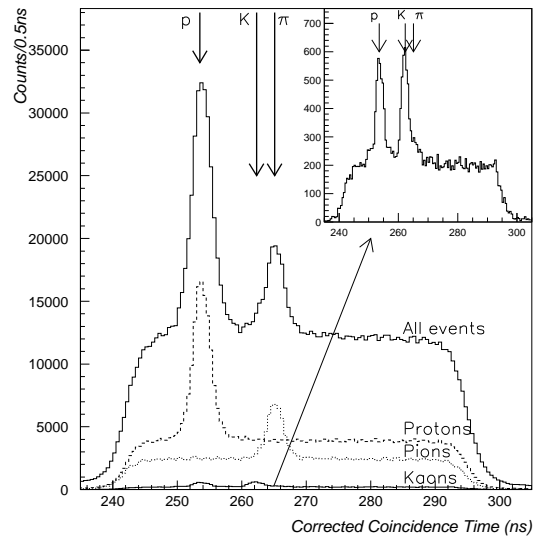}}
\caption{Electron-hadron coincidence time spectra with different 
aerogel-counter based hadron selection.  See the text for further
explanation.}
\label{fig:tcoinc} 
\end{figure}

To perform particle identification (PID) of the knockout kaons, two
new aerogel detectors~\cite{coman00,lagamba,nimHallA}, were built and
installed in the hadron arm to provide kaon identification in the
measured range of momenta, from 1.7 to 2.6 GeV/c.  The counter with
lower refractive index (n=1.015) detected pions, while the higher
index counter (n=1.055) detected both pions and kaons.  Information
from the two detectors was combined to identify pions, kaons and
protons. Figure~\ref{fig:tcoinc} shows the coincidence time spectrum
between the electron and hadron arm.  Without any PID cut the spectrum
is dominated by protons (identified by neither aerogel firing) and
pions (identified by both aerogels firing), kaons amounting only to a
small fraction of the produced hadrons.  The exploded part, showing
kaon events (identified by the first detector not firing and the
second detector firing), shows the kaon peak and the suppression of
the pion peak.  A fraction of protons leak through the selection by
producing Cherenkov light via electron knock-on processes.

\begin{figure}[!ht]
\includegraphics[scale=0.4]{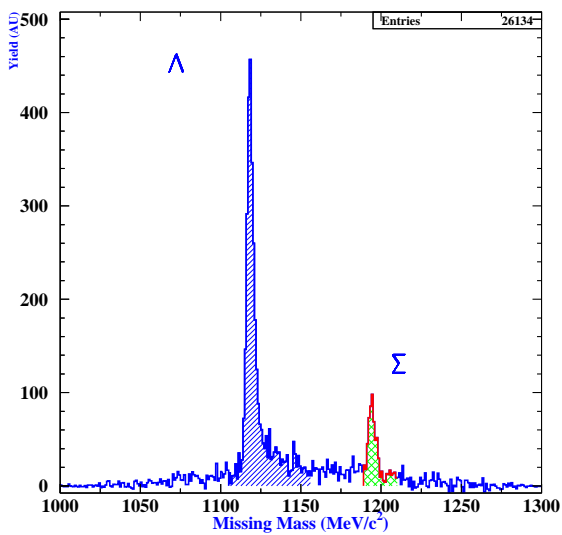}
\caption{\label{missmass} $\Lambda$ and $\Sigma$ missing mass spectra 
obtained at Q$^2$=1.9 GeV$^2$. The region of integration is highlighted.}
\end{figure}

Figure~\ref{missmass} shows the reconstructed mass, $M_x$, of the
unobserved baryon in the $\rm{ ^1H(e,e'K^+)}$ reaction, in this case
either a $\Lambda$ or a $\Sigma^0$ hyperon. Accidental coincidences
were subtracted using a side-band in the timing window. The present
analysis is limited to the exclusive K$^+$--$\Lambda$ production
channel, $e.g.$ $1105.0< M_x< 1155.0$ MeV/c$^2$.
 
Kaon electroproduction was measured at 21 different kinematic
settings.  The ${{d^5\sigma}\over{dE_e' d\Omega_e d\Omega_K}} $ cross
section was determined by comparing the measured yield
(corrected for detector inefficiencies) with the simulated yield from
the Monte Carlo for $(e,e'p)$ reactions (MCEEP) program\cite{mceep},
extended to kaon electroproduction.  MCEEP also was used to determine
the correction for kaon survival (which varied between 15-27\%) and
the radiative correction (which varied between 1.04 and 1.12).  MCEEP
uses a model for kaon electroproduction based on \cite{brauel} updated
as in \cite{cha} to account for cross section variation across the
kinematic acceptance and to extract both the acceptance averaged
and point acceptance $ {{d\sigma(\gamma^*,k)} \over {d\Omega_K}} $.
%
%
%
\begin{table*}[t]
\caption{Unseparated and separated transverse and longitudinal cross 
sections for the $p(e,e'K^+)\Lambda$ reaction as measured in E98-108 
experiment. $E_0$ is the beam energy. Errors are statistical for 
the unseparated and total (statistical and systematical combined in 
quadrature) for separated cross sections.}
\begin{center}
\begin{ruledtabular}
\begin{tabular}{ccccrrr}
\multicolumn{7}{l}{$Q^2 = 1.90$ (GeV/c)$^2$} \\
\hline
\multicolumn{1}{c}{$E_0$} &\multicolumn{1}{c}{$W$}& 
\multicolumn{1}{c}{$t$} &\multicolumn{1}{c}{$\epsilon$}&
\multicolumn{1}{c}{$\sigma_T+\epsilon\,\sigma_L$} 
&\multicolumn{1}{c}{$\sigma_L$}&\multicolumn{1}{c}{$\sigma_T$} 
\\
\multicolumn{1}{c}{[GeV]} & \multicolumn{1}{c}{[GeV]}&
\multicolumn{1}{c}{[(GeV/c)$^2$]} &\multicolumn{1}{c}{}&
\multicolumn{1}{c}{[nb/sr]} &\multicolumn{1}{c}{[nb/sr]}& 
\multicolumn{1}{c}{[nb/sr]}
\\
\hline
5.754&1.91&-0.5994&  0.811&177.6 $\pm$ 8.6   &65.0 $\pm$ 11.6&  125.5 $\pm$ 17.0\\
4.238&        &-0.6183&  0.637&170.7 $\pm$ 11.6 &                             &\\
3.401&        &-0.6183&  0.401&149.7 $\pm$ 11.5 &                             &\\
5.614&1.94&-0.5790&  0.800&178.9 $\pm$ 5.9   &56.4 $\pm$  6.8 &  134.2 $\pm$ 10.5\\
4.238&        &-0.5790&  0.613&170.0 $\pm$ 7.5   &                             &\\
3.401&        &-0.5790&  0.364&154.4 $\pm$ 7.0   &                             &\\
5.754&2.00&-0.5203&  0.792&171.8 $\pm$ 7.1   &80.5 $\pm$ 15.9&  108.0 $\pm$ 23.0\\
4.238&        &-0.5203&  0.575&162.7 $\pm$ 7.4   &                             &  \\
5.614&2.14&-0.4143&  0.726&161.4 $\pm$ 5.4   &36.9 $\pm$ 14.3&  134.6 $\pm$ 21.3\\
4.238&        &-0.4143&  0.471&152.1 $\pm$ 9.4   &                             &\\
\hline
\multicolumn{7}{l}{$Q^2 = 2.35$ (GeV/c)$^2$} \\
\hline
\multicolumn{1}{c}{$E_0$} &\multicolumn{1}{c}{$W$}& 
\multicolumn{1}{c}{$t$} &\multicolumn{1}{c}{$\epsilon$}&
\multicolumn{1}{c}{$\sigma_T+\epsilon\,\sigma_L$} &
\multicolumn{1}{c}{$\sigma_L$}&\multicolumn{1}{c}{$\sigma_T$} 
\\
\multicolumn{1}{c}{[GeV]} & \multicolumn{1}{c}{[GeV]}&
\multicolumn{1}{c}{[(GeV/c)$^2$]} &\multicolumn{1}{c}{}&
\multicolumn{1}{c}{[nb/sr]} &\multicolumn{1}{c}{[nb/sr]}& 
\multicolumn{1}{c}{[nb/sr]}
\\
\hline
5.754&1.80&-0.9498&0.807 &130.1 $\pm$ 6.8  & 53.4 $\pm$ 10.8&  104.3 $\pm$ 17.2\\
5.614&    &-0.9498&0.796 &150.5 $\pm$ 9.7  &                &\\
4.238&    &-0.9498&0.608 &134.7 $\pm$ 7.36 &                &\\
3.401&    &-0.9498&0.359 &130.3 $\pm$ 11.4 &                &\\
5.614&1.85&-0.8562&0.781 &150.1 $\pm$ 6.8  & 64.1 $\pm$ 12.1&   99.3 $\pm$ 18.1\\
4.238&    &-0.8562&0.579 &135.4 $\pm$ 6.4  &                &\\
3.401&    &-0.8562&0.313 &129.1 $\pm$ 10.6 &                &\\
5.614&1.98&-0.6737& 0.737&147.4 $\pm$ 5.8  & 50.9 $\pm$ 12.5&  109.9 $\pm$ 21.4\\
4.238&    &-0.6737&0.494 &135.1 $\pm$ 8.6  &                &\\
5.614&2.08&-0.5716&0.696 &137.2 $\pm$ 4.1  & 68.7 $\pm$  6.8&   89.3 $\pm$ 13.1\\
4.238&    &-0.5716&0.417 &118.1 $\pm$ 6.0  &                &\\
\end{tabular}
\end{ruledtabular}
\end{center}
\end{table*}

The separation of the longitudinal and transverse cross sections was
done using the point cross sections for kaons detected along the
direction of the virtual photon at different values of the virtual
photon polarization parameter $\epsilon$, but keeping $Q^2$, $W$ and
$t$ simultaneously constant (e.g., a Rosenbluth separation).  The
$^1$H$(e,e'K^+)\Lambda$ cross sections are reported in Table I.  The
systematic uncertainties associated with the cross sections are
presented in Table~\ref{systable}.  The total systematic uncertainty
in the cross section amounts to 2.8\%.  Details of the analysis are in
\cite{coman05}.

 \begin{table} 
\caption{\label{systable}Systematic uncertainties for the E981-08 experiment.}
 \begin{tabular}{|c|c|}\hline
Detector/Variable &   Systematic Uncertainty ($\%$)  \\ \hline
Beam energy & 0.12   \\\hline
A1 efficiency  &   0.57   \\\hline
Scintillator efficiencies & 1.33  \\\hline
VDC efficiency &   1.97   \\\hline
A2 efficiency  &   0.87   \\ \hline
Charge  &   0.3 \\ \hline
LH$_2$ target density & 0.2   \\ \hline
Spectrometer acceptance & 0.8 \\\hline
Background subtraction & 0.3 \\ \hline
Kaon absorption & 0.1  \\ \hline
{\bf Quadrature sum} & 2.8 \\ \hline
\end{tabular}
 \end{table}

\begin{figure}
\scalebox{0.4}{\includegraphics[angle=-90]{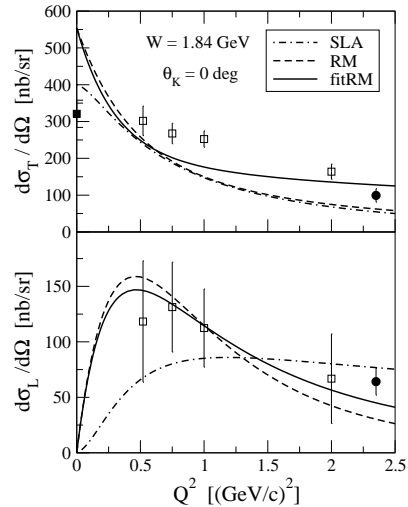}}%
\caption{Separated cross sections from this experiment (full circle) 
and \cite{Moh02} (squares) as a function of $Q^2$ at $W=1.84$ GeV and 
zero kaon scattering angle in comparison with predictions of the 
models (see the text for discussion of the curves).
The photoproduction point (full square) is from 
Ref. \cite{Ble70}.}
\label{fg1}
\end{figure}

%
%
The experiment provides the first good quality separated transverse 
and longitudinal cross sections as a function of energy. In the studied 
kinematical region the cross sections reveal a flat and almost constant 
energy dependence suggesting a quite steep rise of the cross sections 
in the threshold region. The longitudinal cross sections are evidently 
(due to their relatively small error bars) non zero 
in agreement with the rise of $\sigma_{\rm L}$ as a function of energy 
observed by \cite{CLAS} for smaller 
$Q^2$. The data are also consistent with those by \cite{Moh02}, see Fig.~\ref{fg1}.
%
%

\begin{figure*}
\scalebox{0.45}{\includegraphics[angle=-90]{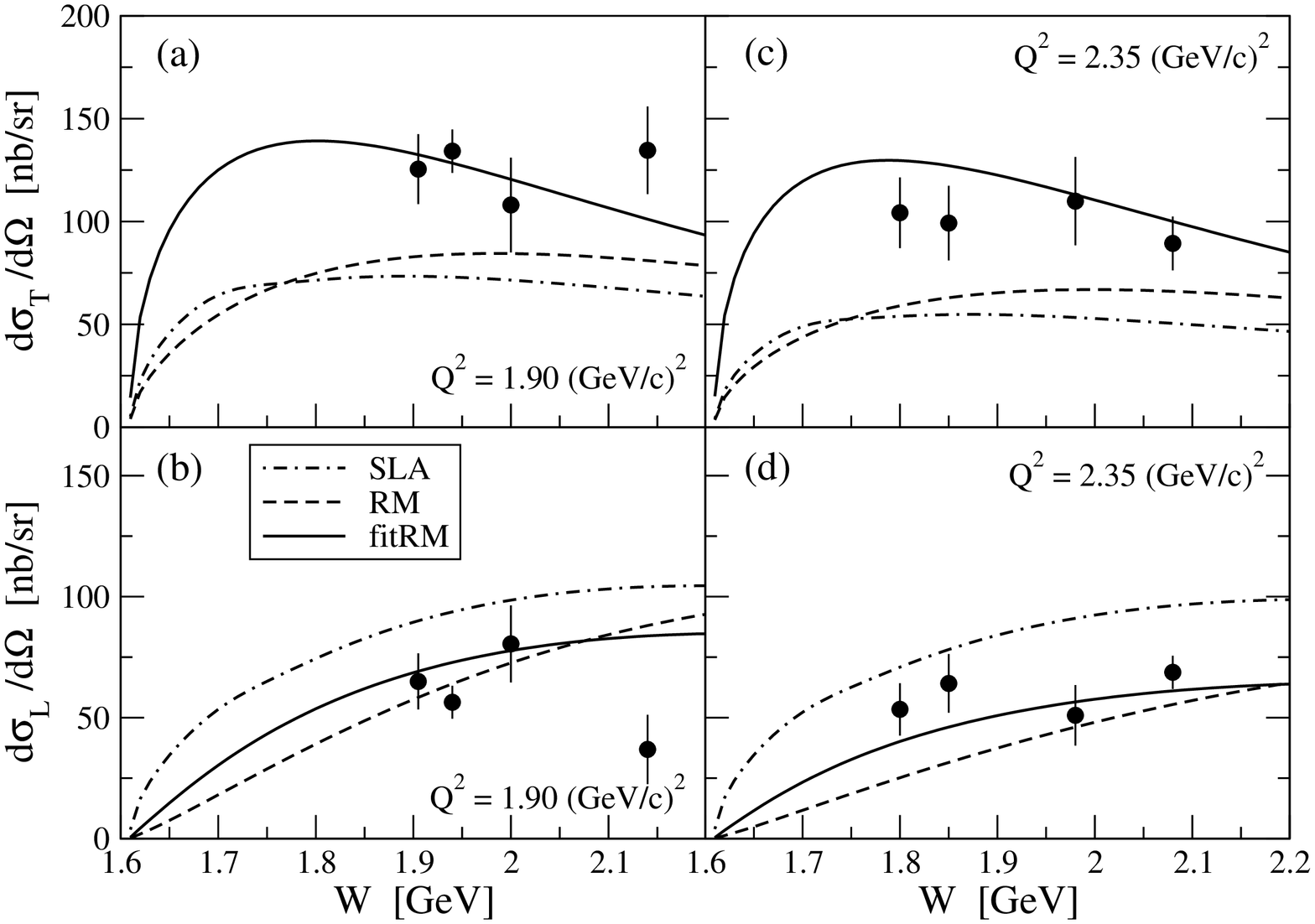}}%
\caption{Energy dependence of separated cross sections from this 
experiment in comparison with predictions of models as in Fig.\ref{fg1} 
for $Q^2=1.90$ GeV, (a) and (b), and 2.35 GeV, (c) and (d), and zero 
kaon angle.}
\label{fg2}
\end{figure*}

The data are shown in Figs.~\ref{fg1} and \ref{fg2} compared to the
isobar Saclay-Lyon (SLA)~\cite{SLA} and Regge (RM)~\cite{RM} model
calculations. Both models systematically underpredict $\sigma_{\rm
  T}$, but give the flat energy dependence observed (Fig.~\ref{fg2}).
The models give different results for $\sigma_{\rm L}$.  The RM model
reproduces the $Q^2$-dependence of $\sigma_{\rm L}$ but not the flat
energy dependence seen in Fig.~\ref{fg2} b and d.  The new data
provide information necessary to refine the models and understand the
dynamics of the process.  A similar inability of the current models to
describe recent data was also shown in Ref.~\cite{CLAS}.

%
%

The flat energy dependence of the cross sections at forward kaon
angles suggests the reaction mechanism is dominated by the non-resonant $t$-channel contribution seen
also in photoproduction \cite{Bra06,Sum06}.  The Regge model was
modified to take into account possible off-mass-shell effects by assuming
a more complex phenomenological prescription for the electromagnetic
form factors of the exchanged K and K* mesons: 
\begin{equation}\label{Fx}
F_x(Q^2,t)= \Lambda_x^2/(\Lambda_x^2+Q^2) + 
              c_x (m_x^2-t)Q^2/(\Lambda_x^2+Q^2)^2
\end{equation}
($x$ = K and K$^*$).  $\Lambda_K^2$ = 0.7 (GeV/c)$^2$ and
$\Lambda_{K*}^2$ = 0.6 (GeV/c)$^2$ were fixed to reproduce mean-square
electromagnetic charge radii of both mesons from previous low $Q^2$
data on the kaon form factor\cite{lowQ}.

The second term in eq. (\ref{Fx}) is the off-mass-shell first-order 
correction in $t$ near the pole.  The prescription
does not violate gauge invariance since in the RM the basic operator
being multiplied by the off-shell form factor is already gauge
invariant \cite{RM}.

The parameters $c_x$ were fit to $c_x$ = 0.87 and 0.79 for K and
K$^*$, from a least square fit to the new data and \cite{Moh02}.
[Note, using only \cite{Moh02} in fitting $c_x$ gives a poor result
showing the impact of the new data.]  Results of the modified Regge
model (fitRM) are shown in Figs.~\ref{fg1} and \ref{fg2} by solid
lines. The off-shell form factors provide both proper normalization
and better energy dependence of the cross sections.  The flatter
$Q^2$-dependence of $\sigma_T$ is described by fitRM, especially for
large $Q^2$, but the known inability of RM to describe photoproduction
cross sections below 3 GeV \cite{Bra06,Sum06} is shown by the
photoproduction point in Fig.~\ref{fg1}.  The ratio
$\sigma_L/\sigma_T$ has the proper decreasing $Q^2$-dependence in
fitRM like RM\cite{RM}, although the ratio depends on the form factors
now.  Improvement in the energy dependence for both $\sigma_T$ and
$\sigma_L$ is also apparent in Fig.~\ref{fg2}.

In summary, new measurements of the longitudinal and transverse
separated cross sections for the $^1$H$(e,e'K^+)\Lambda$ reaction at
Q$^2$ = 1.9 and 2.35 GeV/c$^2$ have been obtained.  These new data 
are in agreement with previous measurements and extend the region for
which the separated cross sections are available.  The data reveal a
flat energy dependence for both longitudinal and transverse cross
sections and put important constraints on models for
kaon electroproduction.

Using the new and previous \cite{Moh02} data on the
electroproduction of kaons we refined the Regge model by including 
phenomenoloigcal off-shell corrections.

Results show the reaction mechanism 
at forward kaon angles and non zero $Q^2$ can be 
described solely by the t-channel exchanges and that the off-shell 
form factors are important to describe the observed flat energy 
dependence, especially for $\sigma_L$. 
The used trajectory form factors, if interpreted as the form 
factors of the K$^+$ and K$^*$ mesons, provide proper values 
of the mean-squared electromagnetic radii and the kaon form 
factor is consistent with the low-$Q^2$ data.

We acknowledge the Jefferson Lab physics and accelerator 
Division staff for the outstanding efforts that made
this work possible. 
This work was supported by U.S. DOE contract 
DE-AC05-84ER40150, Mod. nr. 175,
under which the Southeastern Universities Research Association (SURA)
operates the Thomas Jefferson National Accelerator Facility,
 by the Italian Istituto Nazionale di Fisica 
Nucleare and  by the Grant Agency of the Czech Republic under grant No. 
202/08/0984, and by the U.S. DOE under contracts, DE-AC02-06CH11357,
DE-FG02-99ER41065, and DE-AC02-98-CH10886.

\end{document}